\documentclass[structabstrac]{aa}
\usepackage{amsmath}
\usepackage{amssymb}
\usepackage{mathabx}
\usepackage{graphicx}
\usepackage{natbib}
\usepackage[colorlinks=true,citecolor=blue,linkcolor=red]{hyperref}
\bibpunct{(}{)}{;}{a}{}{,} 
\usepackage{hypcap}
\usepackage[switch,pagewise]{lineno}
\usepackage{subfigure}
\usepackage{overpic}
\usepackage{tablefootnote}
\usepackage{multirow}
\usepackage{verbatim}
\usepackage{url}

\newcommand{\unit}[1]{\ensuremath{\,\mathrm {#1}}}  
\newcommand{\figref}[1]{Fig.~\ref{#1}}
\newcommand{\secref}[1]{Sec.~\ref{#1}}
\newcommand{\avg}[1]{\left< #1 \right>} 
\newcommand{\var}{\ensuremath{\text{Var}}} 
\newcommand{\abs}[1]{\left| #1 \right|} 
\newcommand\alfven{Alfv\'{e}n\xspace}

\newcommand{\dateobsb}{27 Oct 2011\xspace}
\newcommand{\arb}{AR11330\xspace}
\newcommand{\dateobsa}{08 Dec 2010\xspace}
\newcommand{\ara}{AR11131\xspace}

\begin{document}
%
\title{Multi-height observations of magnetoacoustic cut-off frequency in a sunspot atmosphere}

\author{%
Ding Yuan\inst{1,2} \and 
R. Sych\inst{2,3} \and
V.~E. Reznikova\inst{4} \and
V.~M. Nakariakov\inst{1,5,6}
}
\institute{%
Centre for Fusion, Space and Astrophysics, 
Department of Physics, 
University of Warwick,
Coventry CV4 7AL, UK\\
\email{Ding.Yuan@warwick.ac.uk}  \and 
Key Laboratory of Solar Activity, National Astronomical Observatories, Chinese
Academy of Sciences, Beijing, 100012 \and
Institute of Solar-Terrestrial Physics,
Russian Academy of Science, Siberian Branch,
126a Lermontov st.,
 Irkutsk 664033, Russia \and 
 Center for Mathematical Plasma Astrophysics, Department of Mathematics, KU Leuven,
  Celestijnenlaan 200B box 2400 BE-3001 Leuven, Belgium
 \and
School of Space Research, Kyung Hee University, Yongin, 446-701, Gyeonggi, Korea
 \and
Central Astronomical Observatory at
Pulkovo of the Russian Academy of Sciences,
196140 St Petersburg, Russia}
 \date{\today}

\abstract{The cut-off frequency of magnetoacoustic gravity (MAG) waves could be decreased by the inclined magnetic field, and therefore, low-frequency waves could penetrate into the upper atmosphere.}
{We observe the distribution of the cut-off frequency of compressive waves at various heights and reconstruct the magnetic field inclination, according to the MAG wave theory in a stratified atmosphere permeated by a uniform magnetic field.}
{We analysed the emission intensity oscillations of sunspot \ara (\dateobsa) observed at the 1700 \AA{}, 1600 \AA{}, and 304 \AA{} bandpasses of the Atmospheric Imaging Assembly (AIA) onboard the Solar Dynamics Observatory (SDO), and computed the narrow-band power maps with the pixelised wavelet filtering method. The distribution of the cut-off frequency was defined as the median contour in the azimuthally-averaged oscillation power. The magnetic field inclination was estimated with the local cut-off frequency according to the MAG wave theory in the low-$\beta$ limit and was compared to the potential field extrapolation.}
{Shorter period oscillations dominate in the sunspot umbra, while longer period oscillations form an annular shape approximately concentric with the sunspot. Oscillations with longer periods  are distributed further away from the sunspot centre. The 5 min oscillations appear to originate at or lower than the photosphere . The magnetic field inclinations determined with the cut-off frequency theory are about 30-40\% larger than the values obtained by the potential field extrapolation.}
{The oscillation power distribution in a sunspot atmosphere reflects its magnetic and thermal structure. The cut-off frequency could be used to probe the magnetic field inclination, however, other factors have to be included to fully understand this phenomenon. The existence of return magnetic flux at the outer penumbra was evidenced by the cut-off frequency distribution.
}

\titlerunning{Cut-off frequency}
\maketitle
\section{Introduction}
Magnetoacoustic gravity (MAG) waves in the solar atmosphere are subject to strong dispersion:
their properties strongly depend on their frequency. In particular, plasma structuring determines the MAG cut-off frequency.
MAG waves with frequencies below the cut-off value become evanescent and cannot reach the upper layers of the solar atmosphere.
The cut-off value also determines the oscillation frequency of MAG waves excited by a broad-band (i.e., impulsive) source and their propagation to the corona \citep[e.g.,][]{suematsu1982,botha2011,yuan2013fw}. The cut-off frequency of MAG waves in a stratified isothermal atmosphere
permeated by a uniform magnetic field was derived analytically in \citet{bel1977} and \citet{zhugzhda1984}. The MAG cut-off frequency generally depends
on the local plasma $\beta$ and the magnetic field inclination angle $\phi$. In the high-$\beta$ regions (e.g., the photosphere and chromosphere of the quiet Sun), the MAG cut-off frequency reduces to the pure cut-off frequency $\nu_0=\gamma g/4\pi c_\text{s}=5.2\unit{mHz}$, where $\gamma$ is the adiabatic index, $g=274\unit{m/s^2}$ is the gravitational acceleration, and $c_\text{s}=7\unit{km/s}$ is the sound speed in typical chromospheric conditions. In the low-$\beta$ approximation (e.g., in sunspots or active regions), the cut-off frequency is modified  by the magnetic inclination  $\nu_\text{mac}=\nu_0\cos{\phi}$ \citep{bel1977}\footnote{The derivations in \citet{bel1977} contain obvious misprints, but the numerical results and figures are reliable. In this work, we use the trustworthy extreme case at $\beta\ll1$.}. This explains the existence of low-frequency compressive waves in the corona \citep[e.g.,][]{demoortel2012}, which are believed to carry a larger portion of energy flux than their high-frequency counterpart \citep[e.g.,][]{fontenla1993,jefferies2006}. The cut-off frequency of compressive waves could also be used as a seismological tool to probe the local magnetic topology.

According to \citet{bel1977}, the low-frequency waves could penetrate to the upper solar atmosphere along the magnetic waveguides, e.g., sunspots and pores. In particular, due to the inclined magnetic field, the 5 min oscillations, with a period longer than the acoustic cut-off period, could appear in spicules \citep{depontieu2004} and in active region loops  \citep{demoortel2002, depontieu2005}.
Moreover, solar global $p$-modes were observed to reach the
chromosphere because of the inclined magnetic field at the periphery of 
plage regions \citep{dewijn2009}. Long-period oscillations in the corona could also be 
attributed to the channelling effect of the inclined field \citep[e.g.,][]{wang2009, marsh2009,yuan2011}. 

\citet{bloomfield2007} observed the magnetoacoustic cut-off effect in sunspots with
full Stokes (I, U, V, Q) spectropolarimetry and found that the cut-off frequency closely follows the magnetic field modification in the low-$\beta$ case of \citet{bel1977}. However, no further examination was done to compare the information with the magnetic inclination inverted by full Stokes observables. The correlation and phase difference of the time series of the intensity variation obtained at various heights appears more promising after correcting the spatial offsets by considering the magnetic inclination \citep{bloomfield2007}. \citet{mcintosh2006} studied the travel time of narrow-band signals around a sunspot and found good consistency with \citet{bel1977} both in quiet Sun ($\beta\gg1$) and sunspot
($\beta\lesssim1$) cases. \citet{tziotziou2006} applied the empirical formula 
$\nu_\text{peak}(\phi)\approx1.25\nu_\text{mac}(\phi)$ \citep{bogdan2006} to estimate the magnetic field inclination in the sunspot's chromosphere, where $\nu_\text{peak}$ denotes  the frequency of the maximum peak in the power spectrum, and $\nu_\text{mac}$ represents local MAG cut-off frequency modified by the inclined magnetic field.
\citet{reznikova2012a} compared the spectra and phase relations of UV and EUV emission intensity at various heights of the solar
atmosphere and identified the features of upwardly propagating waves.  \citet{reznikova2012a} found that the variation of cut-off frequency across the umbra was consistent with \citet{bel1977}. \citet{reznikova2012b} found a good agreement between the observational magnetoacoustic cut-off frequencies measured in the AIA 304 \AA{} bandpass and the values obtained by the potential field extrapolation \citep{sakurai1982}.

An adiabatic plasma was implicitly assumed in \citet{bel1977}, while radiative losses become significant in the photosphere and chromosphere. The cut-off frequency could also retrieve the plasma parameters associated with radiative losses. \citet{centeno2006,centeno2009} used a linear wave equation with a radiative
cooling term for satisfactory fitting of the observed phase delay and wave amplitude variation
with height in both sunspots and pores. The connectivity of different atmospheric layers determined by the phase
difference and power amplification was demonstrated with multiple spectral line observations \citep{felipe2010}. 

The cut-off frequency of compressive oscillations in a sunspot atmosphere is a good indicator of
local plasma parameters, therefore, it could be used to diagnose the local
atmosphere, i.e., the temperature and magnetic field. In this study, we observed the spatial distribution and height variation of the cut-off frequency with Atmospheric Imaging Assembly \citep[AIA,][]{lemen2011,boerner2011} onboard the Solar Dynamics Observatory (SDO) and estimated the magnetic field
inclination. We present the data sets in \secref{sec:obs}. The methods are
given in \secref{sec:methods}. The statistical and seismological results are summarised in \secref{sec:result}. The conclusions are given in
\secref{sec:con}.

\section{Observations}
\label{sec:obs}
\begin{figure*}[ht]
\centering
\begin{overpic}[width=0.8\textwidth]{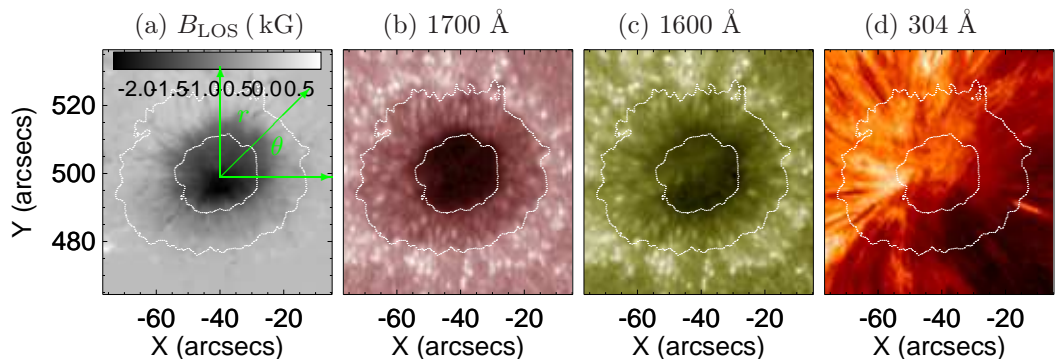}
\put(12,28){ (a) $B_\text{LOS}\left(\unit{kG}\right)$}
\put(20.5,15){\color{green}\vector(1,0){10}}
\put(20.5,15){\color{green}\vector(0,1){10}}
\put(20.5,15){\color{green}\vector(1,1){8}}
\put(25,17){\color{green}$\theta$}
\put(22,20){\color{green}$r$}
\put(35,28){(b) 1700 \AA{}}
\put(56,28){(c) 1600 \AA{}}
\put(78,28){(d) 304 \AA{}}
\end{overpic}
\caption{Multi-instrumental imaging of sunspot \ara (\dateobsa) at different heights.
(a) The line-of-sight magnetic field strength obtained with Helioseismic and Magnetic Imager onboard SDO overlaid with the polar coordinate
system used in this study. (b) - (d): EUV intensity images at different wavelengths obtained with SDO/AIA showing different levels of the sunspot atmosphere. In order of increasing height:  temperature minimum level (1700 \AA{}), upper photosphere and
transition region (1600 \AA{}), and chromosphere (304 \AA{}).
The dotted lines show the sunspot's umbra-penumbra boundary
determined with AIA 4500 \AA{} image.
 \label{fig:aia_fov}}%
\end{figure*}

We selected a large symmetric sunspot associated with active region \ara, where 3 min outwardly propagating EUV disturbances were persistently detectable \citep{reznikova2012a,yuan2012}. The sunspot was situated near the $30^\degree$ latitude in the northern hemisphere. We chose an observation interval without disruptions either from flares nor other transient events. Therefore, the sunspot's magnetic field can be considered as mainly potential. The sunspot crossed the central meridian on \dateobsa. It was least affected by the Wilson depression \citep{longhhead1958} and was well exposed for imaging.  It consisted of a strong magnetic concentration of the south polarity, while the north polarities were spread sparsely to the west and north of the sunspot. The sunspot's shape was almost axially symmetric (\figref{fig:aia_fov}). We used AIA data sets from 02:30 to 03:30 UT at the 1700 \AA{}, 1600 \AA{}, and 304 \AA{} bandpasses. The cadence was 24 s for the 1700 \AA{} and 1600 \AA{} images, and 12 s for 304 \AA{} images. The data sets were prepared with the standard SolarSoft routine \textit{aia\_prep.pro}(v4.11). This sunspot was also analysed in \citet{reznikova2012a, reznikova2012b}.

\section{Methods}
\label{sec:methods}
For each data set $I(x,y,t)$ that showed the time variation of the emission intensity in the plane of sky, we performed
pixelised wavelet filtering (PWF) analysis and obtained narrow-band power maps
$P_\nu(x,y)$ or $P_p(x,y)$, where $p=\nu^{-1}=2, \cdots, 20\unit{min}$
with a resolution of $\Delta p=0.1\unit{min}$.
The wavelet-based PWF method \citep{sych2008,sych2010} is well validated in determining the spatial, temporal, and phase structure of oscillations in an image cube $I(x,y,t)$, where $x$, $y$, and
$t$ are the discrete spatial locations and the sample times, respectively. This method yields the spatial distribution of the power, frequency, and phase of an oscillatory signal, and is intensively used in the study of sunspot oscillations \citep[e.g.,][]{sych2009b}.

\begin{figure}[ht]
\centering
\begin{overpic}[width=0.5\textwidth]{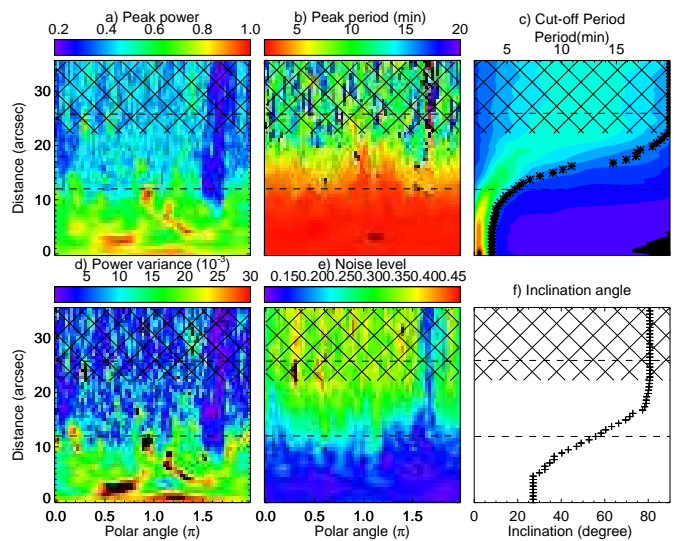}
\end{overpic}
\caption{The 2D distribution of the peak power (a), the peak period (b), the power variance (d), and noise (e) in the 304 \AA{} bandpass.  
The 1D power map as a function of the radius and period (c). The asterisk and plus contours mark the cut-off period (c) and the reconstructed magnetic inclination (f). Panels (a) and (c) have the same colour table. The dashed lines mark the umbra--penumbra boundary. The region of low significance is cross-hatched. \label{fig:vac_304}}%
\end{figure}

\begin{figure}[ht]
\centering
\begin{overpic}[width=0.5\textwidth]{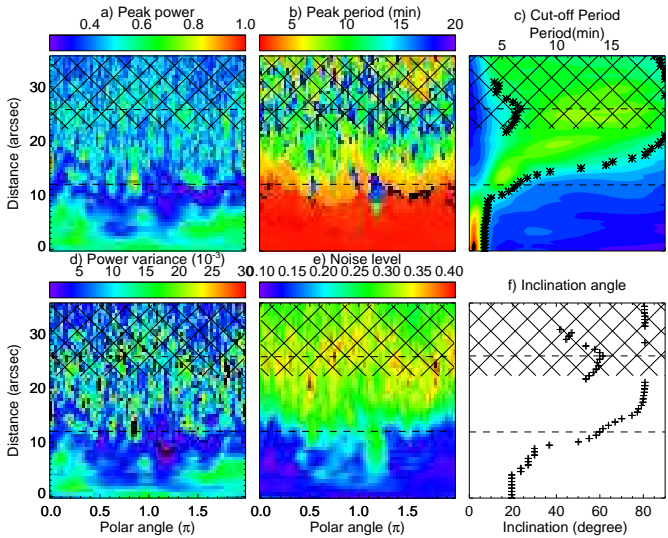}
\end{overpic}
\caption{The same as \figref{fig:vac_304}, but in 1600 \AA{} 
\label{fig:vac_1600}
}%
\end{figure}

\begin{figure}[ht]
\centering
\begin{overpic}[width=0.5\textwidth]{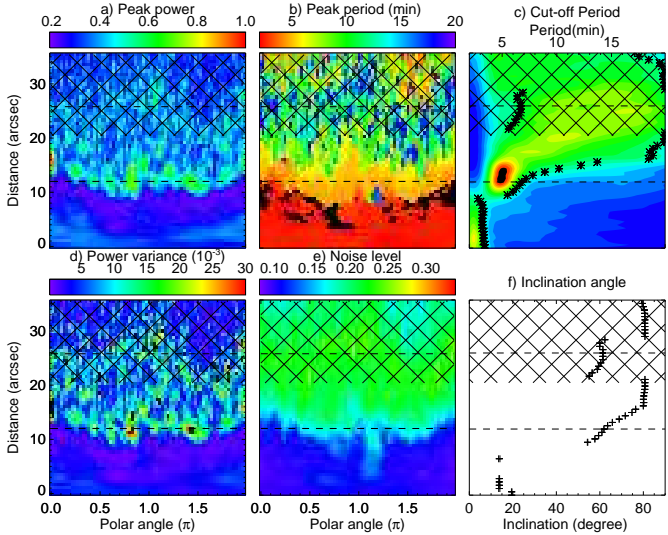}
\end{overpic}
\caption{The same as \figref{fig:vac_304}, but in 1700 \AA{} 
\label{fig:vac_1700}}%
\end{figure}

In sunspots, the spatial distribution of the peak frequencies of compressive perturbations is known to form
a disc (for short periods) or an annulus (for long periods)
concentric at the sunspot centre \citep[e.g.,][]{sych2008,reznikova2012a}. Therefore, to facilitate quantitative
analysis, we transformed the power maps into a polar coordinate $P_\nu(r,\theta)$ with the origin coinciding with
the sunspot centre, where $r$ is the distance to the sunspot centre, and $\theta$ is the
polar angle relative to the horizontal line pointing to the solar west (see \figref{fig:aia_fov} (a)). 
We further averaged the power maps azimuthally to improve the signal-to-noise ratio $P_{\nu}(r)=\avg{P_{\nu}(r,\theta)}_{\theta}$\footnote{The subscripts $\theta$ or $\nu$ denote operations in the spectral or azimuthal dimension, respectively.}, see Fig.~\ref{fig:vac_304}, \ref{fig:vac_1600} and \ref{fig:vac_1700}, panel (c). The cut-off frequency was defined in the global wavelet spectrum as the frequency where the oscillation power decreases to the median level. Likewise, \citet{reznikova2012a} determined the cut-off frequency as the lowest frequency with a power three times above the noise level. Both methods are equivalent, based on the fact that the wavelet spectrum can be approximated by smoothing the corresponding Fourier spectrum and that the noise analyses are directly transferable \citep{torrence1998}. Moreover, our method based on PWF minimises the effects of noise caused by the fine structuring of Fourier spectrum.

We computed the maximum power $P_{\max}(r,\theta)$ (\figref{fig:vac_304}, \ref{fig:vac_1600}, and
\ref{fig:vac_1700}, panels a) and the corresponding peak period $p_\text{peak}(r,\theta)$ (\figref{fig:vac_304}, 
\ref{fig:vac_1600}, and \ref{fig:vac_1700}, panels b). We calculated the variance of the spectral power over the spectrum, $\var(r,\theta)=\var(P_\nu(r,\theta))_\nu$ (\figref{fig:vac_304}, \ref{fig:vac_1600}, and \ref{fig:vac_1700}, panels d). We took the average spectrum $\sigma(r,\theta)=\avg{P_\nu(r,\theta)}_\nu$ to estimate the $1\sigma$ noise level (see \figref{fig:vac_304}, \ref{fig:vac_1600}, and \ref{fig:vac_1700}, panels e). The latter two quantities are good indicators of the significance level in the spectra. The regions of low significance were taken as the regions with the noise level exceeding one standard deviation above the mean noise level, see the cross-hatched regions in  \figref{fig:vac_304}, \ref{fig:vac_1600}, and \ref{fig:vac_1700}. This is equivalent to the region with the power variance less than one standard deviation below the mean variance level. 

\section{Results}
\label{sec:result}
\figref{fig:vac_304}, \ref{fig:vac_1600}, and \ref{fig:vac_1700}, panels a and b illustrate that the oscillation power with periods of less than
$ 3.2\unit{min}$ fills up the umbral region, while those with periods of greater than $3.2\unit{min}$ are normally
distributed in annular structures (or rings) outside the umbra. The oscillations of longer periods form rings of larger radii. We interpret this phenomenon as the modification to the cut-off frequency by the inclined magnetic field \citep{bel1977,depontieu2004,depontieu2005,mcintosh2006}.  \citet{reznikova2012a} demonstrates that there are upwardly propagating compressive waves over this sunspot by the phase analysis. Indeed, the magnetic field lines extending out from the sunspot centre have larger inclination angles. Therefore, the cut-off frequency decreases with increasing distance from the sunspot centre. And hence, at larger distances from the sunspot centre, waves with lower frequencies could penetrate to the upper atmosphere.

\subsection{Spatial and height distribution of dominant oscillations}
\label{sec:result.dist}
The 3 min oscillation power is seen to be strictly constrained within the umbra, and within the discs of radii about 2.1, 5.8, and 7.2 arcsec in the 1700 \AA{}, 1600 \AA{}, and 304 \AA{} bandpasses, respectively (see \figref{fig:vac_304}, \ref{fig:vac_1600}, and \ref{fig:vac_1700}, panels c).
In contrast, the 5 min oscillation power is seen to dominate only at a thin annulus of about 2.9 - 5.0 arcsec in width (see \figref{fig:vac_304}, \ref{fig:vac_1600}, and \ref{fig:vac_1700}, panels c). It is likely that there was strong interaction between the magnetic fields and the $p$-mode seismic waves at that location. \citet{Penn1993} suggested that $p$-mode absorption occurs within a ring surrounding a sunspot umbra. The 5 min oscillation is found to be  strongest in the 1700 \AA{} bandpass at a region where the magnetic field was inclined by 20-50$^\degree$. \citet{cally2003} and \citet{schunker2006} demonstrate that efficient $p$-mode absorption occurs at the height where the \alfven speed reaches the local sound speed and at an attack angle of about $30^\degree$. Therefore, the 5 min oscillation observed near the umbra-penumbra boundary may originate at or lower than the photosphere.

\subsection{Correlation between the cut-off and peak frequencies}
\begin{figure}[ht]
\centering
\begin{overpic}[width=0.4\textwidth]{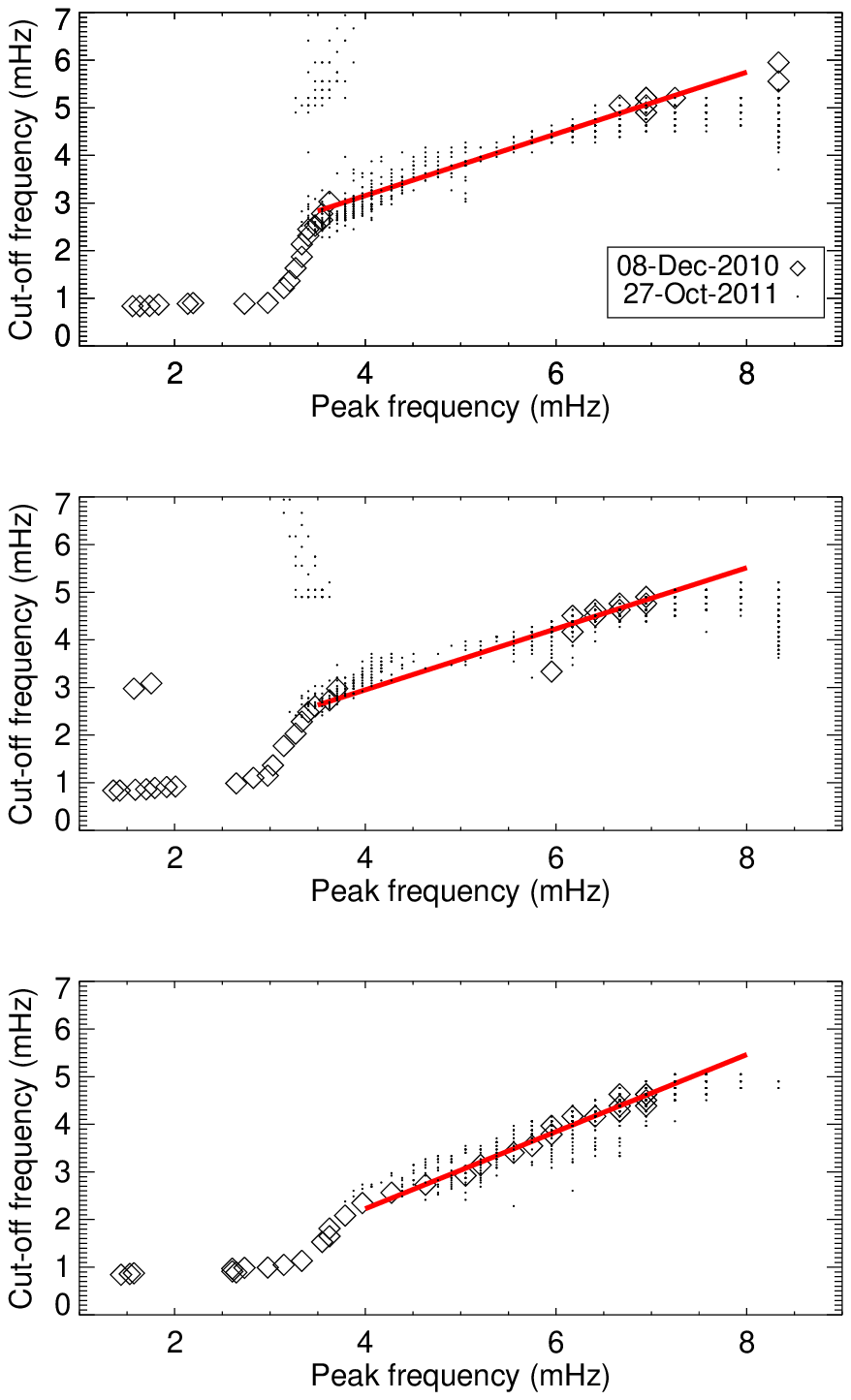}
\put(32,98){\large a) 1700 \AA{}}
\put(14,93){ $\nu_\text{mac}=(0.65\pm0.03)\nu_\text{peak}$}
\put(18,90){ $+(0.57\pm0.18)$}
\put(14,87){ $R=0.996$ }
\put(32,65){\large b) 1600 \AA{}}  
\put(14,60){ $\nu_\text{mac}=(0.64\pm0.05)\nu_\text{peak}$}
\put(18,57){ $+(0.39\pm0.34)$}
\put(14,54){ $R=0.993$ }
\put(32,32){\large c) 304 \AA{}}
\put(14,26){ $\nu_\text{mac}=(0.81\pm0.03)\nu_\text{peak}$}
\put(18,23){$-(1.0\pm0.18)$}
\put(14,20){ $R=0.995$}
\end{overpic}
\caption{Scatter plot and the linear fits (red solid line) between the cut-off frequency and peak
frequency in the 1700 \AA{} (a), 1600 \AA{} (b), and 304 \AA{} (c) data.
The diamond symbols represent the measurements in sunspot \ara (\dateobsa), while dot symbols indicate the data obtained for sunspot \arb 
(\dateobsb).
\label{fig:vac_vpk}}
\end{figure}

We provide a statistical relation between the cut-off and peak frequencies, and therefore, the cut-off frequency (and magnetic inclination angle) could be inferred empirically from a spectrum without referring to complicated analysis.
\figref{fig:vac_vpk} displays the scatter plot of the cut-off frequency $\nu_\text{mac}$ and the peak
frequency $\nu_\text{peak}$, including the measurements made in sunspot \arb (\dateobsb). Below 3.0 mHz, the cut-off frequency $\nu_\text{mac}$ does not vary significantly with the variation of $\nu_\text{peak}$. 
Then, at $\nu_\text{peak}$ between 3.0 mHz and 3.5 - 4.0 mHz, the dependence shows a sharp rise. 
At $\nu_\text{peak}$ greater than 3.5 - 4.0 mHz, a steady increase in $\nu_\text{mac}$ with $\nu_\text{peak}$ is found. 
For the quasi-linear part of the dependence, at $\nu_\text{peak}$ above 3.5 - 4.0 mHz, the Pearson's correlation coefficients are measured at 0.996, 0.993, and 0.995 at the 1700 \AA{}, 1600 \AA{}, and 304 \AA{} bandpasses, respectively. We fitted a linear relationship to the data $\nu_\text{mac}=a\nu_\text{peak}+b$, with constant coefficients $a$ and $b$, using the MPFITEXY routine \footnote{The MPFITEXY 
routine (http://purl.org/mike/mpfitexy) is included in the MPFIT
package \citep[][http://cow.physics.wisc.edu/~craigm/idl/idl.html]{markwardt2009} .} \citep[see][]{williams2010}. The gradients $a$ of the best-fitting linear functions obtained for the 1700 \AA{} and 1600 \AA{} data are very close: $0.65\pm0.03$ and $0.64\pm0.05$, respectively. 
The gradient $a$ in the 304 \AA{} data was $0.81\pm0.03$, in a good agreement with the empirical relation \citep[$\nu_\text{peak}\simeq1.25\nu_\text{mac}$,][]{bogdan2006,tziotziou2006}. However, we obtained non-zero intercepts in the linear fits, $0.57\pm0.18\unit{mHz}$, $0.39\pm0.34\unit{mHz}$ and $-(1.0\pm0.18)\unit{mHz}$ in the 1700 \AA{}, 1600\AA{}, and 304 \AA{} data, respectively (see \figref{fig:vac_vpk}).

 The deviation from the linear relationship in all the analysed channels occurs at the frequency $\nu_\text{peak} \simeq 3.0 - 3.5 \unit{mHz}$ (the 5 min band), see \figref{fig:vac_vpk}. It implies that other physical processes are involved within this frequency range, e.g., energy absorption, as discussed in \secref{sec:result.dist}. The positive non-zero intercepts found in 1700 \AA{} and 1600 \AA{} channels mean that for the value $\nu_\text{peak}$ approaching zero, which is unphysical, the cut-off frequency still exits at the periphery of a sunspot. We speculate that this effect may be caused by a scenario in which the temperature plays a more significant role in the plage region, as suggested in \citet{reznikova2012a}. However, in the 304 \AA{} channel, the intercept is negative, so the isothermal stratified atmosphere model \citep{bel1977} is more realistic at the formation height of the 304 \AA{} bandpass.   

\subsection{Reconstructing the magnetic field inclination}
We reconstructed the magnetic inclination using the simple formula $\theta=\arccos \nu_\text{mac}/\nu_0$ according to the low-$\beta$ case in \citet{bel1977} (\figref{fig:vac_304}, \ref{fig:vac_1600}, and \ref{fig:vac_1700}, panels f). We estimated the magnetic inclination with both the
 theoretical acoustic cut-off frequency $\nu_0=5.2\unit{mHz}$ and the observed maximum cut-off frequency\footnote{The cut-off frequency reached a maximum and remained constant in the inner umbra. This maximum value was taken as the observed maximum cut-off frequency.} $\nu_0^{\lambda}=5.21, 4.90, \text{and}\, 4.63\unit{mHz}$ in the 1700, 1600 and 304 \AA{} channels, respectively. The observed maximum cut-off frequencies are consistent with previous reports \citep[e.g.,][]{worrall2002, mcintosh2006, reznikova2012a,reznikova2012b}. 

\begin{figure}[ht]
\centering
\includegraphics[width=0.3\textwidth]{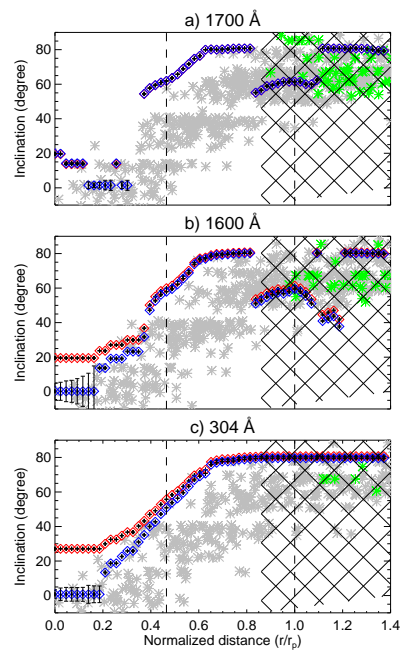}
\caption{Comparison between the field inclination (red and blue diamonds) reconstructed by the cut-off frequency and the values obtained by potential field extrapolation (grey and green asterisks) in the 1700 \AA{} (a), 1600 \AA{} (b), and 340 \AA{} (c) bandpasses. The dashed lines show the umbra-penumbra boundary. The cross-hatched region indicates less-reliable measurements.\label{fig:recon_1d}}%
\end{figure}

  \figref{fig:recon_1d} compares the seismological results and the values obtained with the potential field extrapolation. We followed \citet{reznikova2012a}\footnote{Note that the formation height was implicitly assumed as constant for each observational wavelength.} and retrieved the inclination angles at 500 km (1700 \AA{}), 1100 km (1600 \AA{}), and 2200 km (304 \AA{}) in the potential field. \figref{fig:recon_1d} shows clearly that the general profiles of field inclinations reconstructed
by the cut-off frequency agree well with the potential field extrapolation in the region between 0.2$r_\text{p}$ and 0.8$r_\text{p}$, where $r_\text{p}$ is the average radius of the penumbra. The average offsets of the inclination from the value obtained in the potential field were 32.7/28.8, 27.9/26.6, 30.9/29.7 degrees for $\nu_0/\nu_0^{\lambda}$ reconstructions at the 1700 \AA{}, 1600 \AA{} and 304 \AA{} bandpasses, respectively. The results were slightly improved by using the observed maximum cut-off
frequency (blue diamonds), but the offsets cannot be fully removed.  This implies that the inclined magnetic field is not the only factor that influences the cut-off frequency. It only accounts for about 60-80\% of the decrease in the cut-off frequency.
In the inner umbra (from 0 to 0.2$r_\text{p}$), the cut-off frequency remains almost constant. This suggests that the physics there may be different from the outer part. A region of low inclination is obtained in 0.8$r_\text{p}$ -1.2$r_\text{p}$ in the 1600 \AA{} and
1700 \AA{} bandpasses. In our method, $\cos(\pm\phi)=\abs{\cos(180^\degree\pm\phi)}$
cannot be differentiated \citep[the $180^\degree$ ambiguity
problem, see ][]{metcalf2006}. The low-inclination region in the outer penumbra could be associated with the return-flux with the magnetic vector pointing downward. This could also be connected with sharp structuring of the plasma along the magnetic field. But as our results were obtained on azimuthally-averaged signals integrated for over one hour integration, the latter possibility is less likely.
The return flux in the potential magnetic field was corrected to $[0^\degree,90^\degree]$ (green asterisks). It further confirms our interpretation. Such behaviour is consistent with the return-flux
sunspot model \citep{fla1982, osherovich1982}. The return-flux region was found to spread at about
0.8$r_\text{p}$ - 1.2 $r_\text{p}$. If we assume a parabolic shape of the magnetic field lines, the
returning flux at about 0.4$r_\text{p}$ - 0.6$r_\text{p}$ extends to the height of about 0.14$r_\text{p}$ -
 0.26$r_\text{p}$, and then returns to the surface about 0.4$r_\text{p}$ - 0.6$r_\text{p}$ apart from the
source. The spatial distribution appears to agree with the filamentary reversed polarity fields in the whole penumbra \citep{ruizcobo2013,scharmer2013}.
This effect was less pronounced in the 304 \AA{} data. The agreement between the reconstructed magnetic inclination and the potential field extrapolation was found to be more pronounced in the return-flux region in comparison with the inner sunspot region. 
Indeed, as the return fluxes are normally found at the periphery of the sunspot's
penumbra where the magnetic field is weaker and the temperature is higher, we infer that 
additional effects, e.g., radiative cooling and mode conversion, should be considered to understand the modification of cut-off frequency
there.

\section{Conclusions}
\label{sec:con}
We performed multiple-height observations of the spectral--spatial structure of intensity oscillations in sunspots \ara. Our results confirmed that short-period oscillations are constrained inside the sunspot umbra, while
long-period oscillations were mainly present outside the umbra.
The main spectral peak moves to longer periods with the distance from the sunspot's centre.
The predominant oscillation inside the umbra is the well-known 3 min
oscillation \citep{bogdan2006,demoortel2009}. The 5 min oscillations form a ring shape at the umbra-penumbra boundary. The localisation of the strong power of 5 min oscillations may indicate a strong interaction between the acoustic waves with the magnetic field, where the local condition favoured the absorption of global solar $p$-modes \citep{cally2003, schunker2006}. 

We reconstructed the magnetic field inclination with the MAG wave theory \citep{bel1977}. The results are generally consistent with the potential field extrapolation. The inclined magnetic field was confirmed to decrease the cut-off frequency.  The reconstructed magnetic inclinations are found to be larger than the values obtained in the potential field extrapolation: the magnetic field inclination was estimated to account for about 60-80\% of the decrease in cut-off frequency. This discrepancy suggests that other physics should be included, such as radiative cooling \citep[e.g.,][]{centeno2006,centeno2009,felipe2010} and MHD mode conversion \citep[e.g.,][]{Parchevsky2009,cally2011,Khomenko2012}.  However, the measurement of the cut-off frequency is an alternative and useful method to reconstruct the magnetic field. Empirical dependencies between the dominant oscillation frequency and the cut-off frequency were provided in the observational channels, and hence one could perform simple seismological application with these formulae.

In our method, the assumption $\beta\ll1$ may be invalid in some part of the sunspot. The method could be improved by considering observational or empirical $\beta$ profiles at various heights. If the magnetic field vector field is available as input, the sound speed and therefore local temperature profiles can be reconstructed at various heights. Therefore, the distribution of the cut-off frequency could probe the sunspot's thermal and magnetic structure.

\begin{acknowledgements}
The work was supported  by the RFBR grant N12-02-91161-GFEN\_a, 10-02-00153-a, Marie Curie International Research Staff Exchange Scheme Fellowship  PIRSES-GA-2011-295272, the European Research Council under the \textit{SeismoSun} Research Project No.~321141, the Ministry of Education and Science of the Russian Federation No. 8524, the Russian Foundation of Basic Research under grant 13-02-00044, the Kyung Hee University International Scholarship (VMN), and the Open Research Program KLSA201312 of the Key Laboratory of Solar Activity of the National Astronomical Observatories of China (DY).
The authors acknowledge the anonymous referee's comments for improving the manuscript.
\end{acknowledgements}

\bibliographystyle{aa}
\bibliography{yuan2013aa_cf}
\end{document}